\documentclass[12pt]{revtex4}
\usepackage{textcomp}
\usepackage{setspace}
\usepackage{tipa}
\usepackage{verbatim} 
\usepackage{amsmath}
\usepackage{bbold}
\usepackage{subfigure}
\usepackage{amssymb}
\usepackage{accents}
\usepackage{amscd}
\usepackage{graphicx}
\usepackage[matrix,frame,arrow]{xypic}
\input{Qcircuit}
\input epsf

\newcommand{\be}{\begin{equation}}
\newcommand{\nn}{\nonumber}
\newcommand{\ee}{\end{equation}}
\newcommand{\bea}{\begin{eqnarray}}
\newcommand{\eea}{\end{eqnarray}}

\begin {document}
\title{General U(4) gate for photon polarization and orbital angular momentum}
\author{Craig S. Hamilton$^{1,*}$, Aur\'{e}l G\'{a}bris$^1$, Igor Jex$^1$ and Stephen M. Barnett$^{1,2}$}
\address{$^1$Department of Physics
Faculty of Nuclear Sciences and Physical Engineering,
Czech Technical University in Prague,
B\v{r}ehov\'{a} 7,
Prague,
11519, Czech Republic \\ $^2$ SUPA, Department of Physics, University of Strathclyde, Glasgow G4 0NG, UK} 

\email{craig.hamilton@fjfi.cvut.cz}


\begin{abstract}
We examine the implementation of an arbitrary U(4) gate consisting of CNOT gates and single qubit unitary gates for the Hilbert space of photon spin polarization and two states of photon orbital angular momentum. Our scheme improves over a recently proposed one that uses q-plates because the fidelity is limited only by losses thus in principle it could be used to achieve a perfect transformation. 
\end{abstract}

\maketitle

\section{Introduction}

Photons may have a part to play in quantum information processing (QIP), with polarization having been used so far in various experiments of QIP and quantum cryptography. The downside to photonic qubits is that photon-photon interactions remain a problem as an intermediate medium is required, which generally has very low coupling strengths. A way to circumvent this would be to use more internal states of the photon to carry information, namely the orbital angular momentum states (OAM) \cite{Allen92}.  

Photons carry two types of angular momentum. The first is the well known spin angular momentum, which is the photon's polarization. The second is the OAM of the photons wavefront \cite{Allen92}. The transverse-spatial profile of the wavefront can be decomposed into either Hermite-Gaussian (HG) or Laguerre-Gaussian (LG) polynomials. Both sets of polynomials are characterised by two indices, $m,n$ for HG modes and $l,p$ for LG and the order of the mode is given by $N=m+n=|l|+2p$. The LG set of states have an azimuthal phase dependence of $e^{i l \phi}$ with an orbital angular momentum per photon of $l\hbar$. Recently the OAM states of the photon have began to be manipulated for QIP tasks.

In a recent paper \cite{Sluss09} a device called a `q-box' was constructed and functions as a controlled-U gate in U(4). It uses the Hilbert space of photon polarization and two states of the photons OAM. The latter two states are the Laguerre-Gaussian modes with $l=\pm2$. Although the q-box also uses the states with $l=0,\pm4$ as intermediate steps it finally returns to the $l=\pm2$ space. Up to four of these q-boxes are needed in series to form a general U(4) gate, with wave-plates in-between each q-box. 

However, due to the intrinsic operation of this device, as opposed to losses that could be overcome, it cannot achieve a perfect fidelity. The device operates on the premise that the transverse spatial modes $l=0$ and $l=\pm4$ are separated radially and the wave plates are sized such that they only operate on the $l=0$ mode. However, this mode is not spatially localized within the wave-plate radius (as it would need to be for perfect operation) and thus the device has a reduction in fidelity, as noted in \cite{Sluss09}.

In this paper we describe a method for constructing a general two qubit gate where the qubits are the photon's polarization (H,V) and two states of the photon's OAM. In this scheme we use the OAM states $l=\pm 1,p=0$ (which we label as $\pm$) as our qubit, because these states are a two-dimensional system and as such can be represented on a Poincar\'{e} sphere \cite{Padgett99}. The various single qubit rotations in this OAM Hilbert space can be realised by mode converters \cite{Beijersbergen93} and Dove prisms. The fidelity here is only limited by photon loss and not by any intrinsic limitations. In section~\ref{gatesec} we describe the physical implementation of the single qubit gates and the CNOT gate. In section~\ref{mainsec} we describe from previous literature how to construct a U(4) gate from a series of single qubit gates and CNOTs. Finally we give our conclusions in section~\ref{conc}.

\section{Physical implementation of qubit gates}\label{gatesec}

In this section we describe the actual physical implementation of the gates we use in our realisation of the two qubit gate, which will consist of single qubit gates and a CNOT gate. The spin polarization qubit can be manipulated by half and quarter wave-plates and any single qubit gate, up to an overall phase, can be realised by up to three wave-plates rotated by certain angles about the axis of propagation \cite{Simon90}. 

The OAM qubit can be analogously manipulated by $\pi$ and $\pi/2$ mode converters. A $\pi$ ($\pi/2$) mode converter is a pair of cylindrical lenses separated by a distance of $f$ ($f/\sqrt{2}$), where $f$ is the focal length of the lenses, and it performs the same transformation as a half- (quarter-) wave-plate does on photon polarization, as can be seen from the associated Jones matrices \cite{Allen99}.  Again any single qubit gate can be realised by three mode converters up to an overall phase. The phases we are `ignoring' here can be commuted through a quantum circuit and combined at the end into a single phase shift on the polarization qubit. 

The two qubit CNOT gate can be realised in two different ways. The first way, with the polarization as the control qubit and OAM as the target qubit, is simply a Mach-Zehnder interferometer with a Dove prism in one arm \cite{deOliveira05}. As this requires an interferometer it will be susceptible to phase noise from different path lengths in each arm.

\begin{figure}[phtb]
\begin{center}
\scalebox{0.5}{\includegraphics[trim = 0mm 55mm 40mm 20mm, clip]{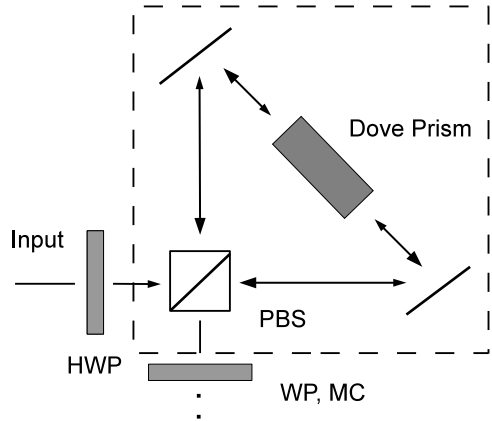}}
\end{center}
\caption{CNOT gate using a Sagnac Interferometer with a half-wave plate before and a series of wave-plates and mode converters following it. PBS- polarizing beam splitter}\label{SI}
\end{figure}

An alternative implementation of the CNOT gate is to use a Sagnac interferometer with a Dove prism in the path. This is equivalent to a CNOT gate (with the OAM qubit as the control and the polarization qubit as the target) up to single qubit gates before the interferometer and following it. These single qubit gates can be absorbed into any gates that are needed in in the implementation of the overall U(4) gate. The Dove prism is rotated at an angle of $\pi/8$ radians and the interferometer uses a polarizing beam-splitter to send the two polarizations along the path in different directions. The matrix representing the action of the Sagnac interferometer in the basis  $ [H+, H-, V+, V-]^T$ is given by,
\be
\hat{U}_{\mbox{\scriptsize{SI}}}=\left [ \begin{array}{cccc}
0 & e^{-il\pi/2} & 0 & 0 \\
e^{-il\pi/2}& 0 & 0&0  \\
0 & 0 & 0 & -e^{-il\pi/2}\\
0 & 0  & -e^{-il\pi/2} & 0  \end{array} \right].
\ee
This gate is shown in figure~\ref{SI}, with the dashed box highlighting the Sagnac interferometer. This setup will be more robust to phase noise than the previous implementation of the CNOT gate as there is only one path that each polarization traverses, thus they acquire the same phase. 

The main source of error in all these gates will be loss from photon reflection at each surface. Components can be made to a high specification to combat this, typically reflections of order $~1\%$ per surface. Our scheme may require many wave-plates/mode converters and thus may have many surfaces contributing, increasing the overall error. For example, the Mach-Zehnder CNOT gate only has a few surfaces and we can therefore expect a fidelity of $>95\%$, although one polarization will experience more loss than the other, due to the Dove prism being in one arm. The other source of errors in this scheme will be phase differences but this can be overcome with careful implementation. 

The gate examined in \cite{Sluss09} was the swap gate and they quoted a fidelity of 83\%, without including errors from photon loss in the components. In our scheme the swap gate would consist of three CNOT gates and four Hadamard gates giving the total number of surfaces to be around 20, which yields an overall loss at $~81\%$ at reflections of 1\%. Better components with reflections of 0.5\% give an overall error of ~90\%. Of course the previous scheme also has to include the figure for reflection loss, making their quoted fidelity lower.   

With the availability of these gates we can implement any U(4) gate with a maximum of three CNOT gates and 12 single qubit unitary gates. In the next section we describe from previous literature how to construct U(4) gate from these elements. Note that depending upon which physical implementation of the CNOT gate we use switches the physical qubits in the quantum circuit diagram and thus the single qubit gate components used. Also it may be advantageous to use both versions of the CNOT gate as it may reduce the number of single qubit gates needed.

\section{Recipe for constructing circuit}\label{mainsec}

The recipe for constructing the U(4) gate relies on the Cartan decomposition of the SU(4) gate into the form \cite{Khaneja01, Kraus01, Tucci},
\[
\Qcircuit @C=1.0em @R=.7em {
& \gate{U_1} & \multigate{1}{e^{-i\hat{H}}} & \gate{V_1} & \qw \\
& \gate{U_2} & \ghost{e^{-i\hat{H}}} & \gate{V_2} & \qw\\
& &\\
& & \mbox{Circuit 1: Cartan decompostion of SU(4)}
}
\]
where $\hat{H} = k_x\hat{\sigma}_{xx} +k_y \hat{\sigma}_{yy}+k_z\hat{\sigma}_{zz}$ and $\hat{\sigma}_{jj}=\hat{\sigma}_j \otimes \hat{\sigma}_j$. Any U(4) gate can be decomposed by adding the appropriate phase $e^{i\theta}$ to make U(4)$\rightarrow$ SU(4) and then compensating for this phase at the end of the circuit. It is then possible to put $\hat{H}$ into the form, which will be used in the next paragraph,
\be
\pi/4 \ge k_x \ge k_y \ge k_z \ge 0, \nn
\ee
using only single qubit gates \cite{Tucci, Zhang03} i.e.~there are classes of U(4) gates which are locally equivalent (different only by single qubit gates).

The operator $e^{-i\hat{H}}$ can be decomposed into, at most, three CNOT gates, with single qubit unitary gates in-between them \cite{Vidal04, Coffey08},
\[
\Qcircuit @C=1.0em @R=.7em {
&\ctrl{1} & \gate{A_1} & \ctrl{1} & \gate{A_2} & \ctrl{1} & \gate{A_3} &\qw \\
&\targ & \gate{B_1} & \targ & \gate{B_2} &\targ & \gate{B_3} &\qw \\
& &\\
&&\mbox{ \hspace{25mm} Circuit 2: Decomposition of  $e^{-i\hat{H}}$}
}
\]
Each $\hat{H}$ can be considered as being in one of four classes, each class representing a necessary number of CNOT gates (0 - 3) to perform the gate $e^{-i\hat{H}}$. Using the ordering of the $k$ parameters above, Vidal and Dawson produced a table for the number of CNOTs needed depending upon the form of $\hat{H}$ (Table I in \cite{Vidal04}). The exact six gates needed are given in \cite{Coffey08}. By combining the two circuit diagrams we can construct a $U(4)$ from a maximum of 3 CNOT gates and 8 single qubit gates.

\section{Conclusions}\label{conc}

We have identified an implementation of a general two-qubit gate, where the qubits are photon polarization and the Laguerre-Gaussian modes of OAM with $l=\pm1$. We used the fact that any U(4) gate can be decomposed into CNOT gates and single qubit unitary gates. Our physical implementation of the gates can, in principle, be error-free and thus a perfect fidelity could be achieved. The main errors in our scheme will be photon reflections at every component surface and its reliance on interferometery, which may pose certain experimental challenges. In the ideal case we can have a perfect fidelity compared to ~80\% fidelity of the scheme of \cite{Sluss09}. One of the interesting applications of our setup is a higher dimensional quantum walk implementation as discussed in \cite{Hamilton10}. However implementations of other QIP schemes can be envisaged. 

\section{Acknowledgements}
C.~S.~H.~ acknowledges financial support from the Doppler Institute in Prague and from grants MSM6840770039 and MSMT LC06002 of the Czech Republic. S.~M.~B. thanks the Royal Society and the Wolfram Foundation for financial support.

\end{document}